\documentstyle[epsf,twoside,fleqn,espcrc2]{article}

\newcommand{\AmS}{{\protect\the\textfont2
  A\kern-.1667em\lower.5ex\hbox{M}\kern-.125emS}}

\def\simm#1{\mathop{\vtop{\ialign{##\crcr
        $\hfil\displaystyle{#1}\hfil$\crcr\noalign{\kern0.5pt\nointerlineskip}
        $\sim$\crcr\noalign{\kern0.5pt}}}}\limits}

\title{
\vspace*{-40pt}
{\normalsize \hfill UTHEP-342} \\
\vspace*{-5pt}
{\normalsize \hfill UTCCP-P-15} \\
\vspace*{-5pt}
{\normalsize \hfill August 1996} \\
\vspace*{-7pt}
Phase structure of QCD for general number of flavors
\thanks{
{Presented by K.\ Kanaya and S.\ Kaya.
at {\it Lattice 96}, St.\ Louis, USA.
}}}

\author{
Y.~Iwasaki\rlap,\address{Center for Computational Physics, 
University of Tsukuba, Ibaraki 305, Japan}%
\mbox{}$^,$\address{Institute of Physics, University of Tsukuba, 
Ibaraki 305, Japan}
K.~Kanaya\rlap,$\mbox{}^{\rm a,b}$
S.~Kaya$\rlap,\mbox{}^{\rm b}$
S.~Sakai\rlap,\address{Faculty of Education, Yamagata University,
                 Yamagata 990, Japan}
and
T.~Yoshi\'e$\mbox{}^{\rm a,b}$
}

\begin{document}
\renewcommand{\textfraction}{0.1}
\renewcommand{\topfraction}{0.9}
\begin{abstract}
We investigate and elucidate the phase structure of QCD for
general number of flavors $N_F$ with Wilson quarks, varying $N_F$ 
from 2 up to 300.
Based on numerical results combined with the result of the 
perturbation theory
we propose the following picture:
When $N_F \ge 17$, there is only
a trivial fixed point and therefore the theory in the continuum limit
is trivial. On the other hand, when $16 \ge N_F \ge 7$, 
there is a non-trivial fixed point and therefore the theory is 
non-trivial with anomalous dimensions, however, without quark confinement.
Theories which satisfy both quark confinement and spontaneous
chiral symmetry breaking in the continuum limit
exist only for $N_F \le 6$.
We also discuss the structure of the deconfining phase 
at finite temperatures
for the small
number of flavors such as $N_F=2$ and 3, through a systematic study
of it for general number of flavors.

\end{abstract}

\maketitle
\setcounter{footnote}{0}
\section{Introduction}
In our previous work \cite{previo}, 
it was shown that in the strong
coupling limit, when the number of flavors $N_F$ is greater than 
or equal to 7,  
quarks are deconfined and chiral
symmetry is restored at zero temperature,
if the quark mass is lighter than a critical
value.
On the other hand, it is well-known that
when $N_F$ exceeds $16 \frac{1}{2}$,
asymptotic freedom is lost. 

In this work we investigate 
the problem of what is
the condition on the number of flavors
for quark confinement and spontaneous chiral symmetry breaking
in the continuum limit,
using the Wilson quark action and the standard 
one-plaquette gauge action. 
We first clarify the phase structure at zero temperature 
for general number of flavors,
in particular, for $N_F \ge 7$. 
When the phase diagram becomes clear, 
we are able to see what kind of continuum limit exists
and eventually answer the problem given above.

In order to reveal the structure of the deconfining phase, 
we increase $N_F$ up to 300,
because the region of the deconfining phase 
at zero temperature
becomes larger with larger $N_F$,
and consequently the structure becomes clear,
as shown below.
Then combining all data including the case of small $N_F$, 
we are able to conjecture the phase structure 
for general number of flavors.
Through this systematic study
of the deconfining phase for general number of flavors,
we are also able to understand
the behavior of the quark mass $m_q$ (defined through an axial-vector
Ward identity) and $m_\pi^2$ as functions of $1/K$
in the deconfining phase at finite temperatures
in particular at small $\beta$
for small number of flavors such as $N_F=2$ and 3.

\section{Simulation parameters}

The lattice sizes are $8^2 \times 10 \times N_t$ 
($N_t =4$, 6 or 8), $16^2 \times 24 \times N_t$ ($N_t=16$)
and $18^2 \times 24 \times N_t$ ($N_t=18$).
We use an anti-periodic boundary condition for quarks 
in the $t$ direction and
periodic boundary conditions otherwise.
When the hadron spectrum is calculated, the lattice is duplicated 
in the direction of lattice size 10 for $N_t \le 8$, 
which we call the $z$ direction.
We call the pion screening mass simply the pion mass and 
similarly for the quark mass.
The current quark mass $m_q$ is defined 
by an axial-vector Ward identity \cite{Bo,ItohNP}.

We use the hybrid R algorithm for the quark matrix inversion.
As $N_F$ increases we have to decrease $\Delta\tau$, such as
$\Delta\tau$ =0.0025 for $N_F=240$, to reduce $O(\Delta\tau^2)$ errors. 
We have checked 
that the errors are sufficiently small selecting typical cases.

\begin{figure}[tb]
\epsfxsize=7cm
\hspace{0.1cm}
\epsfbox{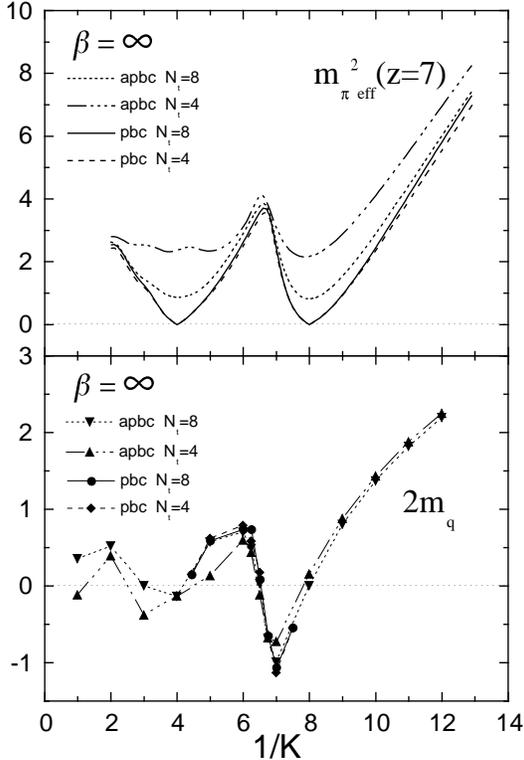}
\vspace{-3.0cm}
\caption{The effective pion mass obtained at z=7 
and the current quark mass for a free quark 
versus $1/K$ 
for $N_t=4$ and $8$ with periodic and anti-periodic 
boundary conditions on $40^3 \times N_t$ lattices. 
}
\label{free}
\vspace{-0.4cm}
\end{figure}

\section{Free case: $\beta=\infty$}
\label{sect:free}
We first investigate the free case,
because the free case is important as a reference to understand
the structure of the deconfining phase.
We will show later that the $1/K$ dependence of the pion mass
in the deconfining phase 
is quite similar to that of free quarks for any $N_F$ and $\beta$.

Because the Wilson formalism lifts the doublers by the Wilson term,
there is only one pole in the free quark propagator at $1/K \simeq 8$.
However, there is reminiscence of doublers. 

In Fig.~\ref{free} we plot the effective pion mass 
obtained at $z=7$ from
the pion propagator.
The four cases shown there correspond to those for
$N_t = 4$ and 8 with an anti-periodic or the periodic boundary condition 
in the $t$ direction. 

In the case of the periodic boundary condition, there are almost no
differences between the $N_t = 4$ and 8 cases. 
The vanishing of $m_q$ at $1/K=8$ and
the behavior of $m_\pi^2$ for $1/K \ge 8$ 
roughly given by $m_\pi^2 \sim (1/K-8)$ 
are all as naively expected.

For $1/K \le 8$, the behavior of $m_\pi^2$ is somewhat complicated.
The vanishing of $m_\pi$ at $1/K =4$ is due to doublers;
one of momenta in spatial direction is $\pi$. 
At $1/K \simeq 6$, 
the momentum-energy (dispersion) relation is completely
opposite to the usual one: The energy is a decreasing function of
the magnitude of momentum for the case of
equal momenta in three spatial dimensions.
The energy becomes infinity at zero momenta like the infinite mass case,
while it is finite for 
other modes of momenta. 
Therefore the appearance of the peak of $m_\pi$ at $1/K \simeq 6$
is due to a superposition of 
contributions from several modes of momenta. 
The behavior of $m_q$ defined through the axial-vector 
Ward identity is also complicated at $1/K \le 8$ due to doublers. 
In the following, we refer to the rightmost $m_q=0$ state
(at $1/K=8$ for $\beta=\infty$)
as simply the massless quark, or $m_q=0$, distinguished from
the other $m_q=0$ states, unless otherwise stated.

When an anti-periodic boundary condition is imposed, the position as well as
the height of the peak of $m_\pi$ at $1/K \simeq 6$ does not change so much.
On the other hand, for the massless quark at $1/K=8$, the pion mass
takes
a value which is roughly twice the lowest Matsubara frequency,
for the both cases of $N_t=4$ and 8.
The behavior around $1/K=4$ is similar, but indicates more complicated 
structure. 
\begin{figure}[tb]
\epsfxsize=7cm
\hspace{0.1cm}
\epsfbox{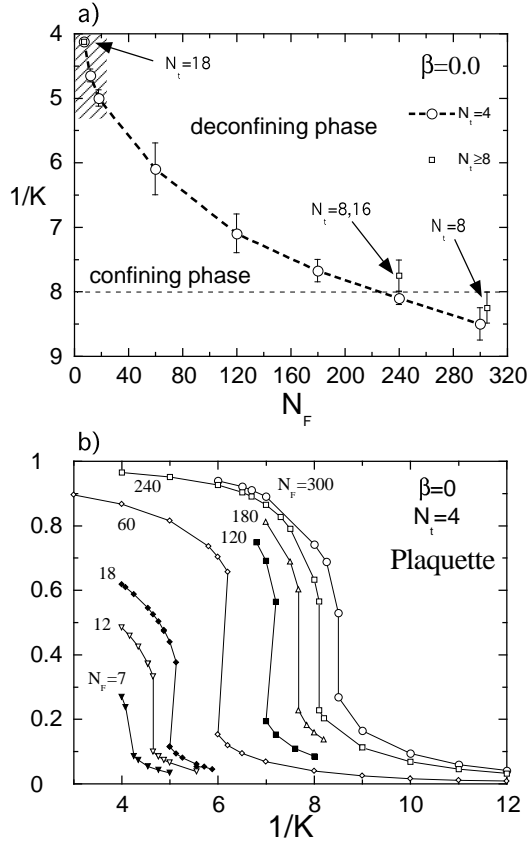}
\vspace{-1.0cm}
\caption{a) The transition point $1/K_d$ at $\beta=0$ 
versus $N_F$ at $N_t=4$ and $N_t\ge 8$. 
For clarity, 
data at $N_t=8$ for $N_F=300$ is slightly shifted in the figure. 
Shaded region was investigated in our previous study 
\protect\cite{previo}.
b) Plaquette versus $1/K$ at $\beta=0$ for various number of $N_F$ at $N_t=4$. }
\label{kd.nf}
\vspace{-0.4cm}
\end{figure}

\begin{figure}[tb]
\epsfxsize=7cm
\hspace{0.2cm}
\epsfbox{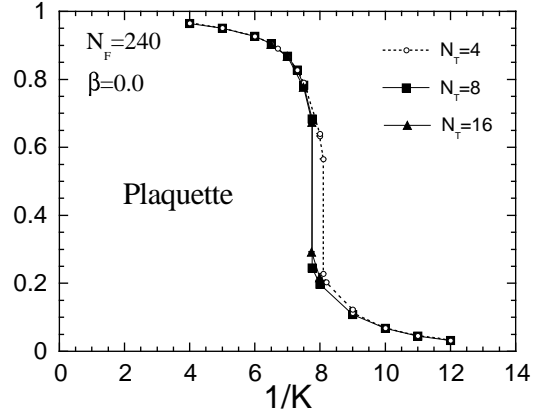}
\vspace{-1.0cm}
\caption{Plaquette at $\beta=0$ for $N_F=240$ at $N_t=4$, 8, and 16.}
\label{nf240.pl}
\vspace{-0.4cm}
\end{figure}

\section{$K_d$ versus $N_F$}
We now study the case of $\beta=0$ as in the previous work \cite{previo}.
Fig.~\ref{kd.nf}a shows the results of the deconfining transition 
points $K_d$ at $\beta=0$ obtained on $N_t=4$ lattices 
for various numbers of $N_F$
 (see also Fig.~\ref{kd.nf}b).

The data of plaquette (Fig.~\ref{nf240.pl}) for $N_F=240$
at $\beta=0$ 
indicates that the deconfining transition is of first order 
and that the location of the transition point 
is independent of $N_t$ for large $N_t$ 
($1/K_d \simeq 8.1$ for $N_t=4$ and $1/K_d \simeq 7.75$ for $N_t=8$ 
and 16).
Therefore we conclude that the transition is bulk as is confirmed 
in our previous work for $N_F=7$ \cite{previo}.
In Fig.~\ref{kd.nf}a the transition points at $N_t\ge 8$ 
for $N_F=7$, 240 and 300 are also included.  
These values are roughly those for the bulk transition
points at zero temperature.

As $N_F$ increases up to 240, $1/K_d$ approaches toward or exceeds
the value 8, which is the value of $1/K$ for a massless free quark.
Because of this, for $N_F \simm{>} 240$, we are able to see 
a wide region of the deconfining phase 
in the whole range of $\beta$.
Therefore we first
intensively investigate the case $N_F=240$, 
and then decrease $N_F$.

\section{$N_F=240$}

\begin{figure}[t]
\epsfxsize=7cm
\hspace{0.3cm}
\epsfbox{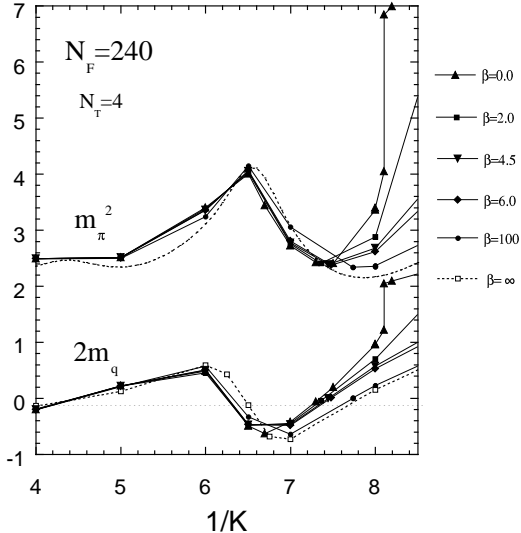}
\vspace{-1.0cm}
\caption{$m_\pi^2$ and 2$m_q$ versus $1/K$ for $N_F=240$ at $N_t=4$.}
\label{nf240.mass}
\vspace{-0.4cm}
\end{figure}

Fig.~\ref{nf240.mass}
shows the results of $m_\pi^2$ and $2m_q$ for $N_F=240$ 
at $\beta=0$, 2.0, 4.5, 6.0, and 100 on the $N_t=4$ lattice.
A very striking fact is that the shape of $m_\pi^2$ and $2m_q$ 
as a function of $1/K$
only slightly changes for $1/K < 8$
when the value of $\beta$ decreases from $\infty$
down 0. 
Only the position of the local minimum of $m_\pi^2$ at $1/K\simeq8$, 
which corresponds to the vanishing point of $m_q$, 
slightly shifts toward smaller $1/K$.
We obtain similar results also for $N_t=8$.

From these data we propose the phase diagram
in Fig.~\ref{nf240.phase} for $N_F=240$.
The dark shaded line is the phase boundary between the 
confining phase and the deconfining phase at zero temperature.
When $N_t =4$ or 8, the boundary line bends down at finite $\beta$
as shown in Fig.~\ref{nf240.phase}, 
due to the finite temperature phase transition
of the confining phase.

The dashed line corresponds to the $m_q=0$ line. 
This line also corresponds to the minimum point of $m_\pi^2$.
We have also calculated the quark propagator in the Landau gauge and
checked that  chiral symmetry of the propagator,
$\gamma_5 G(z) \gamma_5 = - G(z)$,
is actually satisfied on the $m_q=0$ line. 

The results for the case $N_F=300$ are essentially the same as those
for $N_F=240$ except for very small shifts of the transition point
and the minimum point of $m_\pi^2$.

\section{Renormalization group flow}

\begin{figure}[tb]
\epsfxsize=7cm
\epsfbox{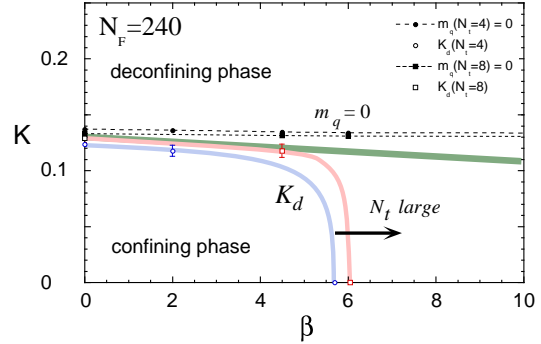}
\vspace{-1.0cm}
\caption{Phase diagram for $N_F=240$.}
\label{nf240.phase}
\vspace{-0.4cm}
\end{figure}

The $m_q=0$ point at $\beta=\infty$ is the trivial IR
fixed point for $N_F \ge 17$. 
The phase diagram shown in Fig.~\ref{nf240.phase} suggests that 
there are no other fixed points on the $m_q=0$ line at finite $\beta$.

In order to confirm this, we investigate 
the direction of Renormalization Group (RG) flow along the $m_q=0$ line
for $N_F=240$, 
using a Monte Carlo Renormalization Group (MCRG) method.

One problem here is that it is practically impossible
to make simulations in the massless limit at zero temperature 
due to the existence of zero modes in the quark matrix.
Therefore, we impose an anti-periodic boundary condition 
in the $t$ direction.

We make a block transformation for a change of scale factor 2,
and estimate the quantity $\Delta \beta$ for the change of 
$a \rightarrow a'=2a$.
We generate configurations
on an $8^4$ lattice on the $m_q=0$ points at $\beta=0$ and 6.0
and make twice blockings. 
We also generate configurations
on a $4^4$ lattice 
and make once a blocking.
Then we calculate $\Delta\beta$ 
by matching the value of the plaquette at each step.\footnote{
It is known for the pure $SU(3)$ gauge theory, 
in particular in the deconfining
phase, that one has to make a more careful analysis 
using several types of Wilson loop
to extract a precise value of $\Delta \beta$.
We reserve elaboration of this point and a fine tuning of $1/K$ at each
$\beta$ for future works.
}

From the matching, we obtain $\Delta \beta \simeq 6.5$ for $\beta=0$ 
and $10.5$ for $\beta=6.0$.
The value obtained from the perturbation theory is 
$\Delta \beta \simeq 8.8$ at $\beta = 6.0$ for $N_F=240$.
The signs are the same and the magnitudes are comparable.
This suggests that the directions of RG flow on the $m_q=0$ line
at $\beta=0$ and 6.0 are the same as that at $\beta=\infty$ for $N_F=240$.
This further suggests that there are no fixed points at finite $\beta$.
All of the above imply that the theory is trivial in the case of $N_F=240$.

\section{$240 \ge N_F \ge 17$}

\begin{figure}[tb]
\epsfxsize=7cm
\hspace{0.1cm}
\epsfbox{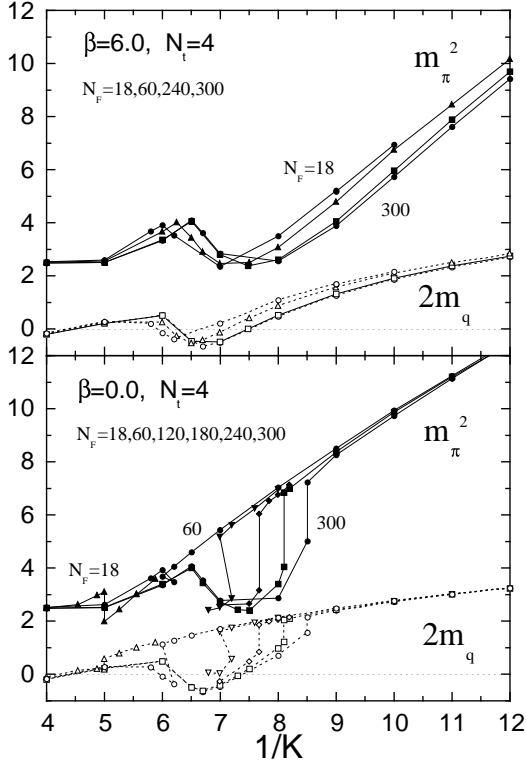}
\vspace{-1.0cm}
\caption{$m_\pi^2$ and 2$m_q$ 
for $N_F \le 18$ ,$N_t=4$ at $\beta=$6.0 and 0.0 versus $1/K$.}
\label{300to18}
\vspace{-0.4cm}
\end{figure}

Now we decrease the value of $N_F$ from 240.
Fig.~\ref{300to18} shows $m_\pi^2$ and $2m_q$ 
for various $N_F$, from 300 down to 18,
at $\beta=6.0$ and 0.

When $\beta=6.0$, the shapes of $m_\pi^2$ are almost identical to each other,
except for a slight shift toward smaller $1/K$ as $N_F$ decreases.

When $\beta=0$, the boundary of the first order
phase transition between the deconfining phase and the confining phase
moves toward smaller $1/K$ and therefore the range of
the deconfining phase decreases: $1/K_d \simeq 8.5$, 8, 7.6, 7.2, 6.1 for
$N_F= 300$, 240, 180, 120 and 60, respectively, 
as shown in Fig.~\ref{kd.nf}.
For $N_F=18$, the value of $1/K_d$ decreases down to 5.0. 

\begin{figure}[tb]
\epsfxsize=7cm
\epsfbox{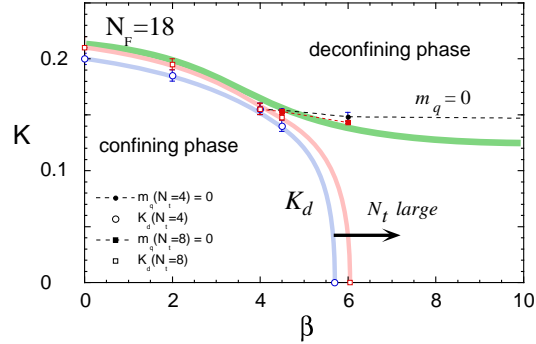}
\vspace{-1.0cm}
\caption{Phase diagram for $N_F=18$.}
\label{nf18.phase}
\vspace{-0.4cm}
\end{figure}

Due to the fact that the confining
phase invades the deconfining phase, the massless line in the
deconfining phase hits the boundary at finite $\beta$ when $N_F$
becomes small. 
For example, in the case of $N_F=18$, 
as shown in Fig.~\ref{nf18.phase}, 
it hits at $\beta = 4.0$ --- 4.5.

Although the area of the deconfining phase decreases with 
decreasing $N_F$, 
the shape  and the position
of $m_\pi^2$ in the part of deconfining phase
only slightly change from $N_F=300$ to 18.
The values of $m_q$ as functions of $1/K$ 
also show only slight changes in the deconfining phase. 
These facts, combined with the perturbative result
that for $N_F \ge 17$, $\beta=\infty$ is the IR fixed
point, suggest that the RG flow along the massless quark in the
deconfining phase is the same as that for $N_F=240$.
In this case, the theory is trivial for $N_F \ge 17$.%
\footnote{
Although we have investigated only down to the $N_F=18$ case, 
we do not expect
any qualitative differences between the cases of $N_F=17$ and 18.
}

\section{$16 \ge N_F \ge 7$}

When $N_F \le 16$, the theory is asymptotic free. 
On the other hand, 
quark confinement is lost for
$N_F \ge 7$ even in the strong coupling limit $\beta=0$ \cite{previo}.
Therefore the question is what happens 
for the cases $16 \ge N_F \ge 7$ in the deconfining phase.

\begin{figure}[t]
\epsfxsize=6cm
\hspace{0.5cm}
\epsfbox{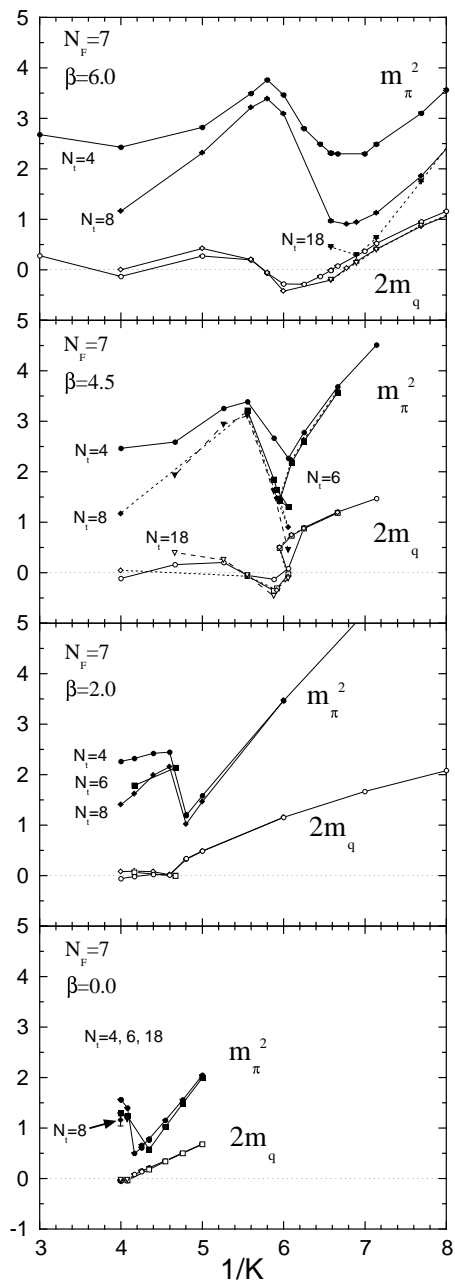}
\vspace{-1.0cm}
\caption{$m_\pi^2$ and 2$m_q$ for $N_F=7$ versus $1/K$.}
\label{nf7}
\vspace{-0.4cm}
\end{figure}

We have intensively simulated the cases $N_F=12$ and 7.
See the results for the case $N_F=7$ in  Fig.~\ref{nf7}.
The gross 
feature of the phase diagram does not much differ from the case of
$N_F=18$ which is shown in Fig.~\ref{nf18.phase}.

One crucial point is that there is no chiral limit in the confining
phase at least for $\beta \simm{<} 4.5$. 
This suggests that quarks are not confined in the continuum limit.

A more subtle point is whether there is a difference between the case
$N_F \ge 17$ and the case $16 \ge N_F \ge 7$ concerning the deconfining
phase.
For $N_F \le 16$, the massless point at $\beta=\infty$ is a UV
fixed point. 
Therefore the direction of the RG flow around
$\beta=\infty$ is opposite to that for $N_F \ge 17$.
In order to study
the phase structure around $\beta =0$,
we have performed a MCRG study for the case $N_F=12$ similar to
that for the case of $N_F=240$.
We have investigated the RG flow along the maximum of $m_\pi^2$,
because this line roughly corresponds to an infinite quark mass line
in a sense explained in Sec.~\ref{sect:free}. We have found that the RG flow
at $\beta=0$ and $\beta=6.0$ are opposite and the flow changes the direction
around $\beta=4.0$.
This result suggests that there is a fixed point on the $m_q=0$ line.
The flow of RG on the $m_q=0$ line
at $\beta=5.0$ is the same as that at $\beta=\infty$,
which is opposite to that of the case of $N_F=240$.
This implies that the location of the fixed point is at $\beta <  5.0$.
Because of the fact that
the massless line $m_q=0$ hits the boundary between the two phases
around $\beta=4.0$,
it is difficult to determine the location of the fixed point.
Although we have not been able to locate the fixed point,
we propose the phase structure,
based on the consideration given above,
that
there is a non-trivial IR fixed point at finite $\beta$ \cite{Banks1}
for $16 \ge N_F \ge 7$.
Of course we have to make a detailed RG study to confirm
that there is a non-trivial fixed point,
which we reserve for future works.
The existence of a non-trivial fixed point implies 
that the theory in the continuum limit is a 
non-trivial theory with anomalous dimensions, 
however, without confinement. 

\section{$N_F \le 6$}

For $N_F = 2 $ and 3 at finite temperatures,
it was shown that $m_q$ 
has an unexpected $1/K$ dependence in the deconfining phase
when $\beta$ is smaller than about 5.3 \cite{milc1,milc2,LAT94}. 
The $m_\pi^2$ as a function of $1/K$ also shows a cusp.
In this section we concentrate on these issues.
(For the discussion on the phase structure at zero temperature,
see Ref.~\cite{stand} and references in it.)

\begin{figure}[tb]
\epsfxsize=6.9cm
\hspace{0.2cm}
\epsfbox{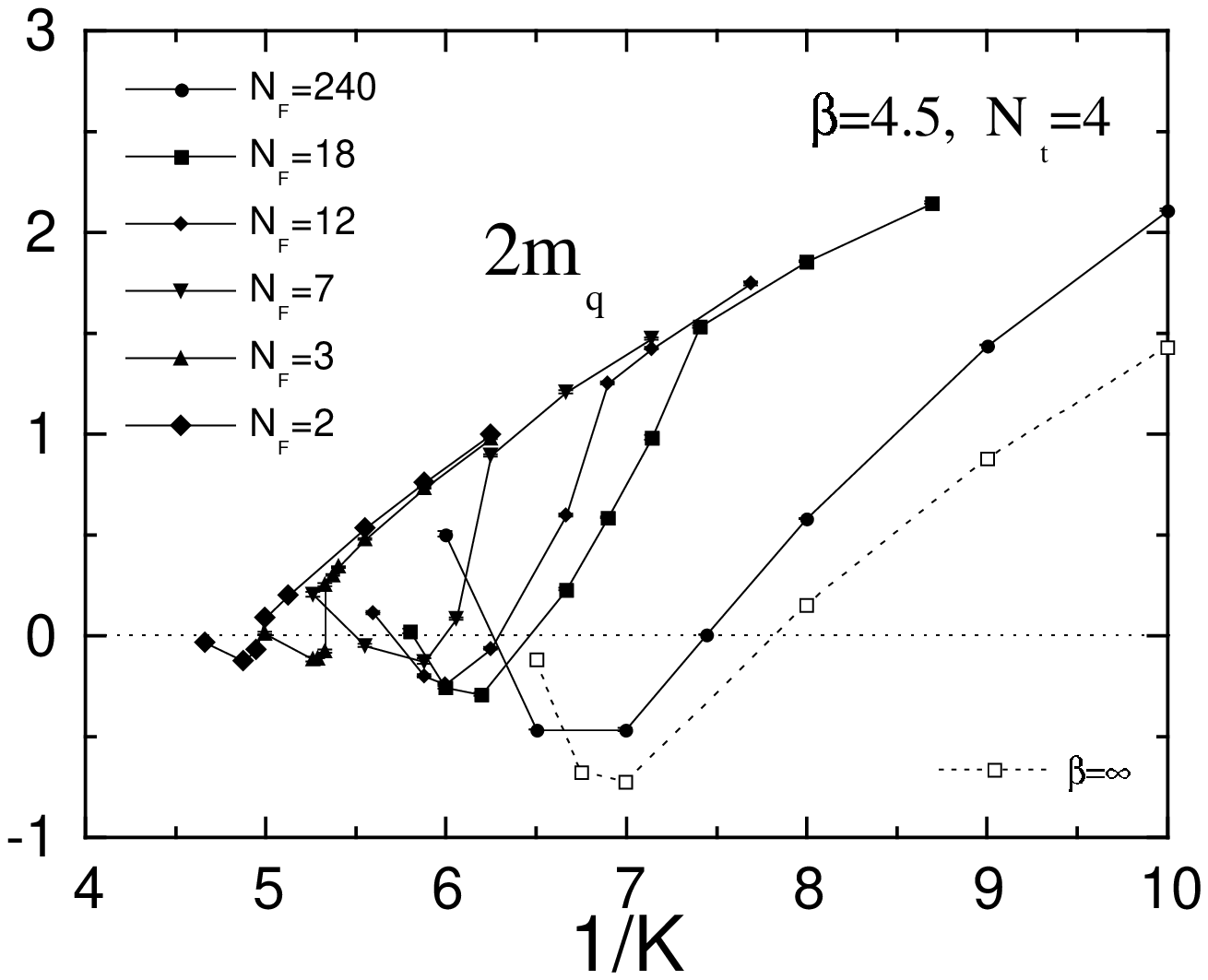}
\vspace{-1.0cm}
\caption{$2m_q$ for various number of flavors at $\beta=4.5$ 
versus $1/K$.}
\label{nf2.mq}
\vspace{-0.4cm}
\end{figure}

We point out
that the behaviors of $m_q$ and $m_\pi$ in the high temperature 
deconfining phase for $N_F \le 6$ at $\beta \simm{<} 5.3$
are similar to those in the 
deconfining phase at zero temperature for larger $N_F$. 
In Fig.~\ref{nf2.mq}, the value of $2m_q$ is plotted vs.\ $1/K$
for from $N_F=240$ down to $N_F=2$ at $\beta=4.5$. 
Results at $\beta=5.0$ are similar.
The value of $2m_q$ for the free quark case is also plotted. 
There are two major trends when $N_F$ decreases from 240 down to 2:
One is an overall shift of the function $2m_q$ in
the deconfining phase to smaller $1/K$ and shrinkage of the shape
with decreasing $N_F$. The other is a shift of the transition point
from the confining phase to the deconfining phase toward smaller $1/K$.

It should be noted that
except for the overall shift and shrinkage mentioned above,
the shape of the function $2m_q$ in terms of $1/K$ 
in the deconfining phase 
is quite similar for all $N_F$ and also similar to that of a free quark.
Thus the behavior of $m_q$ as well as $m_\pi^2$ 
as functions of $1/K$ in the deconfining phase at $\beta \simm{<} 5.3$
can be understood essentially as those of free quarks for $1/K \le 8$.

\begin{figure}[t]
\epsfxsize=6.6cm
\hspace{0.1cm}
\epsfbox{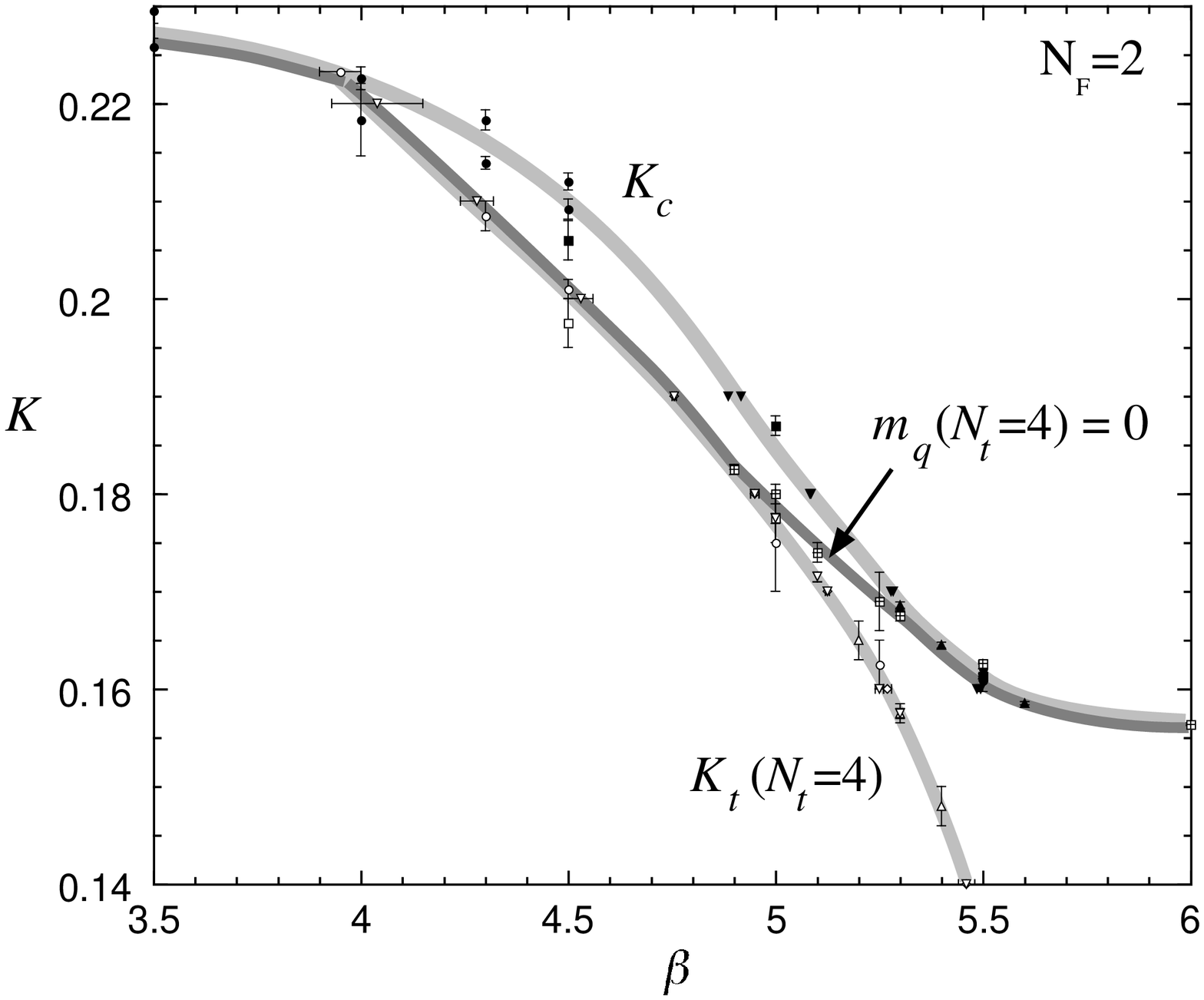}
\vspace{-1.0cm}
\caption{Phase diagram for $N_F=2$.}
\label{nf2.phase}
\vspace{-0.4cm}
\end{figure}

We further note that for $N_F=2$ and 3, the massless quark
point (the rightmost $m_q=0$) in the deconfining phase is 
different from the $K_c$ point
in the confining phase for $\beta \simm{<} 5.3$.
The difference 
is illustrated
in the phase diagram for $N_F=2$ on the $(\beta,K)$ plane
(see Fig.~\ref{nf2.phase}).
On the $K_c$ line, both $m_\pi^2$ and $m_q$ vanish 
in the confining phase at zero temperature \cite{stand}.
The finite temperature transition for $N_t=4$ occur on the line
$K_t$. 
The massless quark line for $N_t=4$ 
in the high temperature deconfining phase
agrees with the $K_c$ line at $\beta \simeq \infty$.
However, they deviate from each other at $\beta \simm{<} 5.3$.
This corresponds to the fact that 
the value of $m_q$ does not depend on the phase when $\beta \simm{>} 5.5$,
while it strongly depends on the phase 
when $\beta \simm{<} 5.3$ \cite{stand,rev}.

\section{Conclusions}
Based on numerical results combined with the result of 
the perturbation theory
we propose the following picture:
There are three categories depending on the number
of flavors $N_F$: a free theory for $N_F \ge 17$, a non-trivial IR
fixed point for $16 \ge N_F \ge 7$ and confinement and spontaneous
chiral symmetry breaking for $N_F \le 6$.
The conclusion that there are three categories depending on $N_F$
is in accord with recent theoretical works of
$N=1$ super-symmetric QCD \cite{Seib}.

We also find that, for any $N_F \ge 7$ and for $N_F \le 6$ at 
$\beta \simm{<} 5.3$,
the $1/K$ dependences of $m_q$ and $m_\pi^2$
in the deconfining phase are essentially the same
as those of free quarks for $1/K \le 8$,
which show non-trivial structure due to doublers. 
The peculiar behavior of $m_q$ as well as the cusp structure
of $m_\pi^2$ in the deconfining phase at $\beta \simm{<} 5.3$ 
for $N_F=2$ \cite{milc1,milc2,LAT94,rev}
can be understood 
in terms of 
the overall shift in $1/K$ and shrinkage of this structure.

Numerical simulations are performed with 
HITAC S820/80 at KEK, and Fujitsu VPP500/30 and
QCDPAX at the University of Tsukuba.
This work is in part supported by 
the Grants-in-Aid of Ministry of Education,
Science and Culture (Nos.07NP0401, 07640375 and 07640376).

\end{document}